\documentclass{ws-procs9x6}
\begin{document}

\title{Measurement of the analyzing power in pp elastic scattering 
in the peak CNI region at RHIC}

\author{
\underline{H.~Okada$^1$}, 
I.G.~Alekseev$^2$, A.~Bravar$^3$, 
G.~Bunce$^{3,4}$, S.~Dhawan$^5$, R.~Gill$^3$,  
W.~Haeberli$^6$,  O.~Jinnouchi$^4$,  
A.~Khodinov$^7$, A.~Kponou$^3$,   
 Y.~Makdisi$^3$, W.~Meng$^3$, A.~Nass$^3$, N.~Saito$^{1,9}$, H.~Spinka$^{10}$, 
E.J.~Stephenson$^{11}$, D.N.~Svirida$^2$,  
 T.~Wise$^6$, A.~Zelenski$^3$}


\address{
  ${}^1$Kyoto Univ.~
  ${}^2$ITEP~
  ${}^3$BNL~
  ${}^4$RIKEN BNL Research Center~ 
  ${}^5$Yale Univ.~
  ${}^6$Univ. of Wisconsin-Madison ~
  ${}^7$Stony Brook Univ.~
  ${}^8$RIKEN~
  ${}^{9}$ANL~ 
  ${}^{10}$Indiana Univ.\\
  E-mail: okadah@bnl.gov
}

\maketitle

\abstracts{
The analyzing power $A_N$ for pp elastic scattering is expected to reach a peak 
value of $0.045$ in the Coulomb Nuclear Interference (CNI) region at a momentum
transfer $-t$ of $0.003$ $(\rm{GeV/}\textit{c})^2$. During the $2004$ RHIC Run, we 
completed a measurement of $A_N$ in the CNI region by detecting the recoil protons 
from \textit{pp} elastic scattering using a polarized atomic hydrogen gas jet target and the
$100$ GeV RHIC proton beam.
We report the first measurements of  the $A_N$ absolute value and
shape in the $-t$ range from $0.0015$ to $0.010$ $(\rm{GeV/\textit{c})^2}$ with a precision
better than $0.005$ for each $A_N$ data point. The recoil protons were detected with
two arrays of Si detectors. The absolute target polarization as monitored by a 
Breit-Rabi polarimeter was stable at $0.924\pm 0.018$.
This result allows us to further investigate the spin dependence of elastic
pp scattering in the very low $-t$ region. } 


The elastic scattering of nucleons has been studied for a long time as the most 
fundamental reaction to extract information on the elementary building blocks of 
matter, nucleons. Regarding the initial and final states in elastic scattering 
of protons, there are three types of processes ; no spin reversal, single spin reversal
and double spin reversal processes. These processes can be described by the use of five 
independent helicity amplitudes which have hadronic and electromagnetic parts.
The electromagnetic part of the helicity amplitudes is well understood by QED. On the other 
hand, the hadronic part is being studied\cite{c1}.
The analyzing power ($A_N$) mainly depends on the single spin reversal process. 
At very small momentum transfer $-t$ = $0.001$ to $0.02$ $(\rm{GeV/}\textit{c})^2$, 
two helicity amplitudes (hadronic and electro magnetic) become similar in size and 
interfere with each other. We call this phenomenon Coulomb Nuclear Interference (CNI).
The values of $A_N$ is expected to reach a peak of $0.045$ at $-t \sim0.003$ $(\rm{GeV/}\textit{c})^2$. 
The shape of $A_N$ is related to the hadronic amplitude of the single spin reversal process. 
Experimetal knowledge is needed to constrain theory. The only $A_N$ data available in the CNI 
region is from E$704$ at Fermilab\cite{c2} with limited accuracy. Last spring (the $2004$ RHIC run), 
we carried out a measurement of $A_N$ in the peak CNI region using a polarized atomic hydrogen gas jet target. 

Since the CNI process is ideal to be used as a polarimeter for a high energy proton beam\cite{c3} 
and to determine the RHIC beam polarization, we have installed the RHIC Absolute Polarimeter\cite{c4}. 
This polarimeter consists of the silicon spectrometer and the polarized atomic hydrogen-jet target, 
which includes a Breit-Rabi polarimeter to obtain a precise absolute target polarization\cite{c5}. 
By knowing the absolute polarization of the target ($P_{target}$), we measured the absolute $A_N$ from 
the scattering asymmetry with respect to the target spin sign ($\epsilon_{target}$), that is, 
$A_N$= $\epsilon_{target}$/$P_{target}$. 
Since we are using the elastic scattering process of proton beam and proton target, we also obtain
the absolute beam polarization from the scattering asymmetry with respect to the beam and $A_N$.
Figure $1$ shows the schematic of the setup. We detect the recoil particles by using 3 left-right pairs 
of silicon strip detectors. The strips run vertically with respect to the beam direction. 
The read out pitch is about $4$ mm and each channel outputs energy ($T_R$) and time of flight (\textit{tof}).
The readout channel number gives us the angle ($\theta_R$) information as shown in Fig. 2.

\begin{figure}[htbp]
\begin{tabular}{cc}
\begin{minipage}{0.50\hsize}
\begin{center}
\includegraphics[width=\hsize]{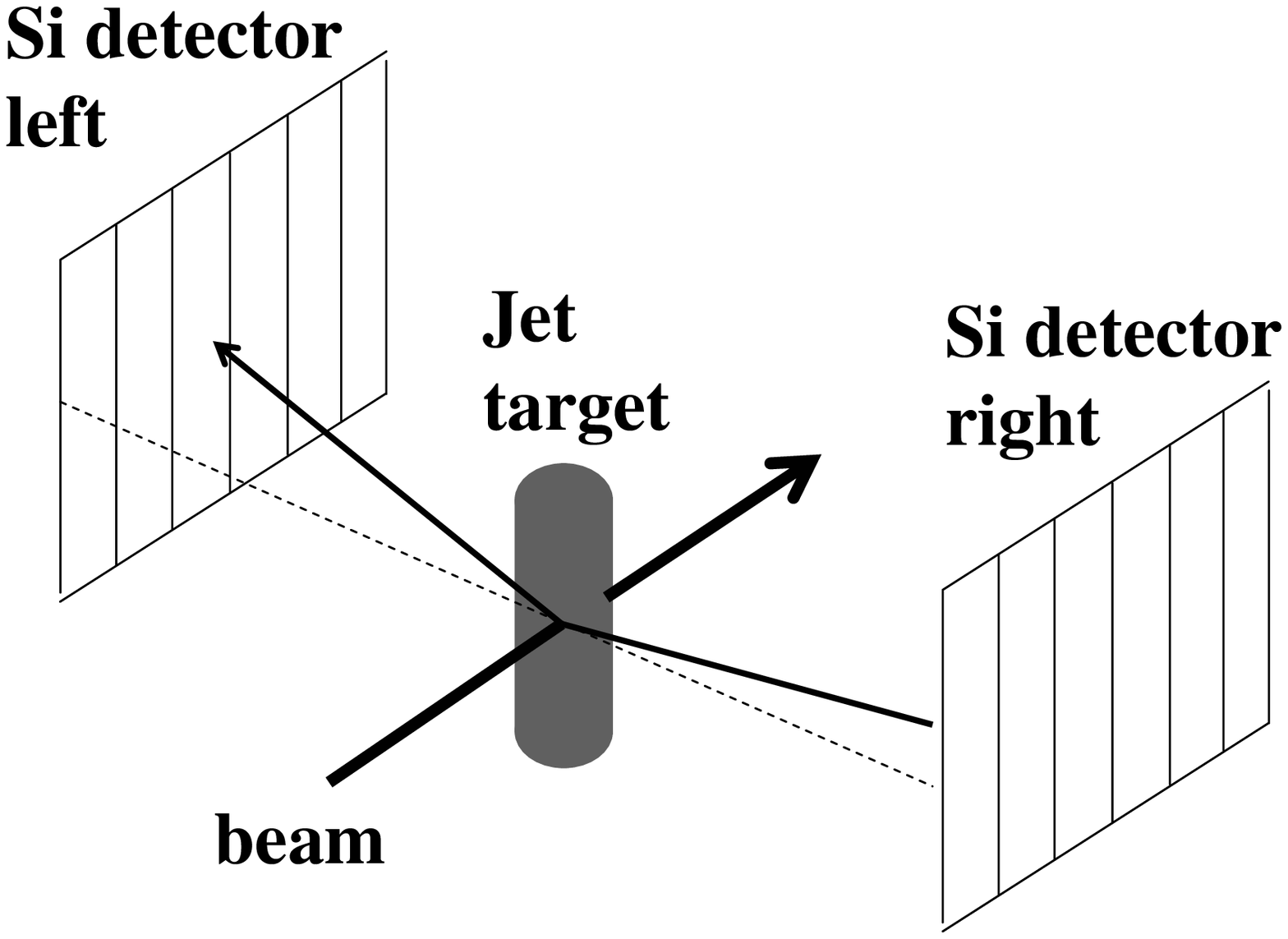}
\caption{Setup}
\label{Setup}
\end{center}
\end{minipage}
\begin{minipage}{0.40\hsize}
\begin{center}
\includegraphics[width=\hsize]{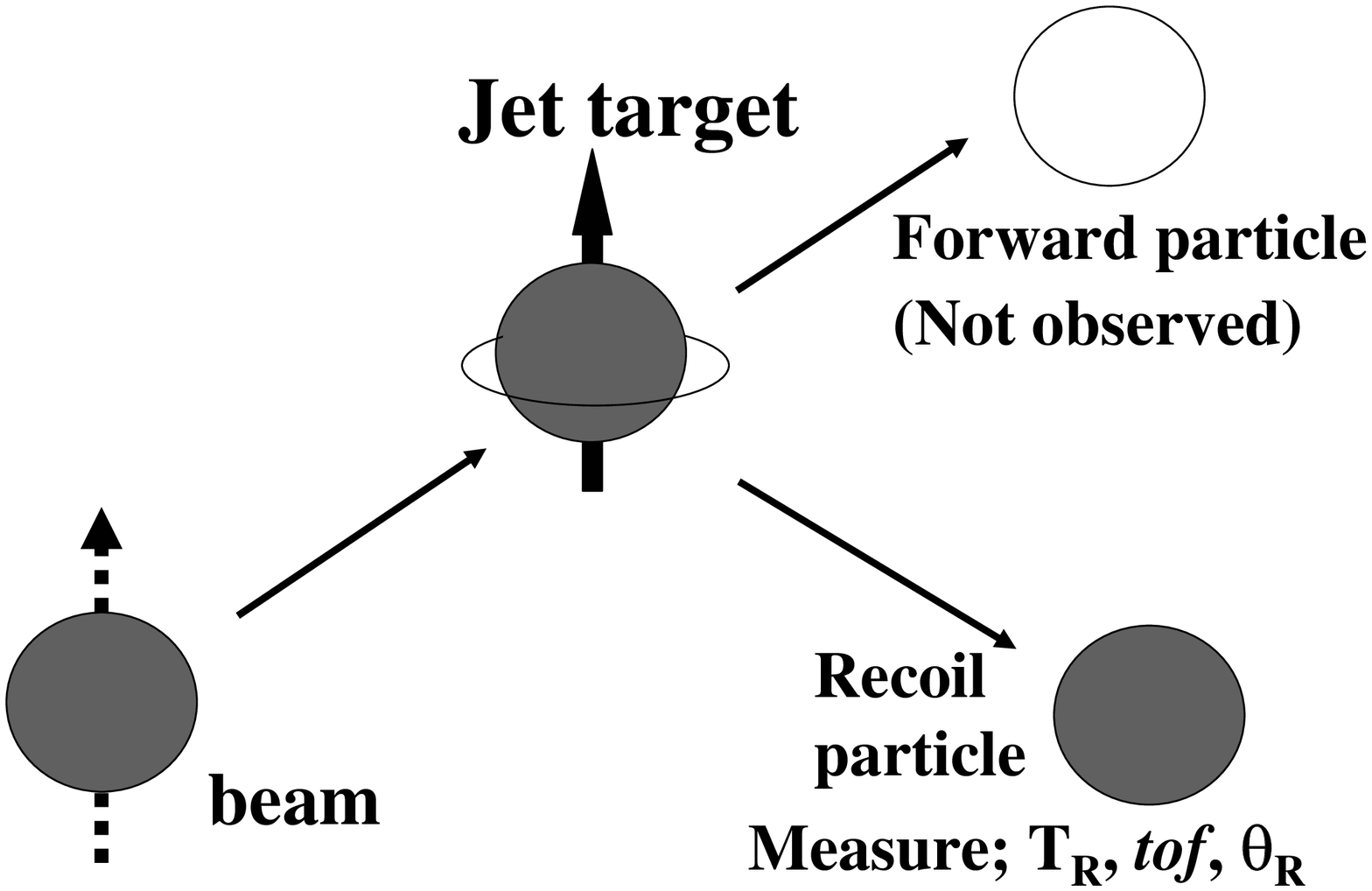}
\caption{Elastic scattering process}
\label{Setup}
\end{center}
\end{minipage}
\end{tabular}
\end{figure}

To get $A_N$, we measure the left-right asymmetry of the elastic event yield with respect the target
spin sign, then divide it by target polarization. The key points of this experiment are 
high target polarization and the capability of selecting the elastic events. The target 
polarization was quite high, accurate ($P_{target}$=$0.924\pm 0.018$) and stable\cite{c5}. 
We confirmed that both the recoil particle mass and the forward-scattered 
particle mass (missing mass) were consistent with that of the proton using the following two correlations. 
$T_R$ and \textit{tof} confirm the recoil particle is proton (Fig. $3$).
$T_R$ and channel number ($\varpropto \theta_R$) confirm the forward scattered particle is proton (Fig. $4$). 
We have focused on the smaller $T_R$ region (\verb|<| $5$ MeV) which corresponds to $-t$ region of 
$0.0015<-t < 0.01{\rm(GeV/c)^2} $.

\begin{figure}[htbp]
\begin{center}
\includegraphics[width=0.5\linewidth]{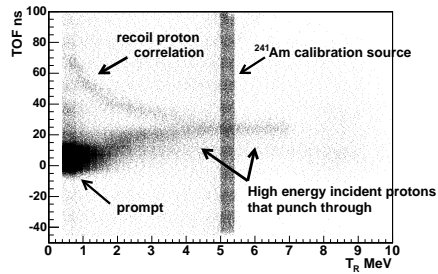}
\caption{The correlation of $T_R$ and \textit{tof} of one detector. This confirms the recoil particle is a proton.}
\end{center}
\end{figure}
 
\begin{figure}[htb]
\begin{center}
\includegraphics[width=0.5\linewidth]{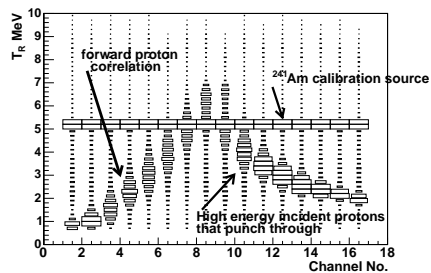}
\caption{The correlation of $T_R$ and channel number $(\varpropto \theta_R)$ of one detector. Each detector 
has $16$ channels. This confirms the forward scattered particle is a proton (less than $5$ MeV). 
Because the higher energy protons punched through the silicon 
detector, the measured $T_R$ decreases beyond channel number $9$.}
\end{center}
\end{figure}

Figure $5$ shows our result (closed circles) and E$704$ data (open circles). 
The errors are statistical only. We estimate the systematic error in $A_N$ to be 
$0.0015$, which is dominated by possible contributions of background (beam gas interaction etc.). 
The fraction of background is less than $5$\%. 

\begin{figure}[htbp]
\begin{center}
\includegraphics[width=0.7\linewidth]{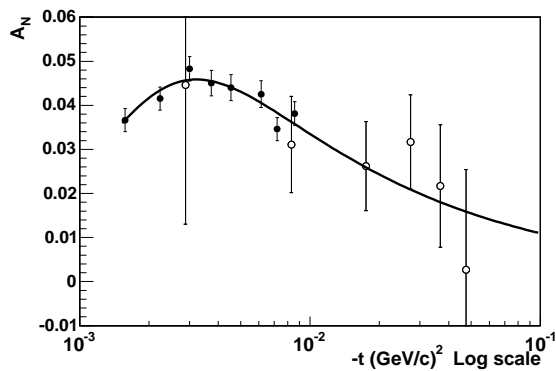}
\caption{$A_N$ data for \textit{pp} elastic scattering as a function of $-t$.
The closed $8$ circles are the result; the open circles are from $E704$, 
The errors shown are statistical only. 
The solid line is a no hadronic spin flip calculation. $\chi^2$/d.o.f.= $6/8$, 
which includs statistical and systematic errors. }
\end{center}
\end{figure}

We have measured $A_N$ in the peak CNI region from $0.0015$ to $0.010$ $(\rm{GeV}/\textit{c})^2$ in $-t$,
and observed the peak shape of $A_N$ for the first time.  
\\
\\
The authors would like to thank the BNL Instrumental Division for their work on the silicon detectors 
and electronics. The work is performed under the auspices of U.S. DOE contract Nos. DE-AC02-98CH10886 
and W-31-109-ENG-38, DOE grant No. DE-FG02-88ER40438, NSF grant PHY-0100348, and with support from RIKEN, Japan.


\begin{thebibliography}{0}
\bibitem{c1} T.L. Trueman, these proceedings (2004).

\bibitem{c2} N. Akchurin \textit{et al}.,Phys. Rev. D \textbf{48}, 3026 (1993).

\bibitem{c3} O. Jinnouchi \textit{et al}., these proceedings (2004) and AIP Conf. Proc. \textbf{675} (2003) 817.

\bibitem{c4} A. Bravar \textit{et al}., these proceedings (2004) and AIP Conf. Proc. \textbf{675} (2003) 830.

\bibitem{c5} A. Zelenski, T. Wise and A. Nass \textit{et al}., these proceedings (2004).

\end{thebibliography}
\end{document}